\documentclass[12pt]{article}

\usepackage{}

\begin{document}

\title{\bf Hadron Diffraction: Results and Problems\footnote{A capsule version of my talk at the 19th International Seminar "QUARKS-2016"}}

\author{V.A.~Petrov \\
Institute for High Energy Physics,\\
``NRC Kurchatov Institute'', Protvino, RF}

\date{}

\maketitle

\begin{abstract}
This is a cursory review of diffractive studies ( mostly elastic scattering)at the LHC with physical interpretation
of the experimental data and comments to related problems of theory.
\end{abstract}

\section{Introduction}
What is hadron diffraction? We mean under this term a class of processes in hadron-hadron collisions which are characterized by specific angular distribution of the scattered particles similar to diffraction patterns in optics and by the mating feature: large rapidity gaps. The particles to be detected are produced at small angles w.r.t. the colliding beams.
Why are such studies worthy of being made? Which fundamental problems of modern physics can be resolved with help of the corresponding data?
In terms of characteristic spatio-temporal scales diffractive studies deal with
distances and times much larger than those which characterize the majority of the LHC experiments, e.g. the Higgs or hard jets production. Moreover, this difference
gets more drastic with the energy growth. When we say "larger" we mean "larger or even significantly larger than 1 fermi". Well, such distances are a routine feature
in low-energy nuclear physics.Why expend energy to rediscover them again?
The matter is that 1 fm at 100 MeV grossly differs from 1 fm at 1 TeV. In what follows we try to show this exposing the latest results obtained in diffractive studies at the LHC and trying to understand their physical meaning.

\section{A Bit of History}

One of the most claimed quantities in high-energy
physics is total cross-section $ \sigma_{\mathrm{tot}} $ as function
of energy, $\sqrt{s} = 2\sqrt{p^{2}+ m^{2}} $ in the center-of-mass
frame ($ p $ is the c.m.s momentum of the colliding protons and $ m
$ is the proton mass). This quantity is often loosely described as
"an effective area" (transverse to the beam(s) direction) where
something happens with colliding hadrons. The formal definition is
\begin{equation}
\sigma_{\mathrm{tot}} = \sum\limits_X\sigma_X ,
\end{equation}
where $\sigma_X$ is the cross-section of any possible process $
p+p\longrightarrow X $. "Optical theorem" states that the total
cross section is simply related with the imaginary part of the
amplitude $ T(s,t) $ of elastic scattering $ p+p\longrightarrow p+p $\,:
\begin{equation}
\sigma_{\mathrm{tot}}\cdot 2p\sqrt{s} = \lim_{\theta\rightarrow 0}
Im T(s,t),
\end{equation}
where $t = -2p^{2}(1-\cos \theta) \approx - p^{2}\theta^{2}$. We
carefully use the symbol $ \textit{lim} $ to stress that zero-angle
scattering doesn't mean that "nothing happened". With no massless
exchange particle a non-trivial limit exists. The optical theorem
shows that the total cross-section is not only a geometrical
quantity but is also defined by the intensity of interaction
embodied in the amplitude. We also see that the total cross-section
which is, on the one hand, an integral characteristic, is, on the other hand,
related to the local value of the scattering amplitude (at zero angle)
and therefore is relevant to diffraction.
 Some 50 years ago physicists believed
that at very high energies total cross-sections achieve a limiting
constant value naturally defined by a "strong interaction radius" $
\sim m_{\pi}^{-1} $. The only dissident was W. Heisenberg who
predicted $ \sigma_{\mathrm{tot}} \sim \ln^2\!s $. However his
derivation was discarded by fellow theoreticians like L.D. Landau \textit{et alii}
for years to come. When in 1971 experiments at the 70 GeV "Serpukhov"
proton synchrotron in Protvino showed that the cross-section of
interaction of the $  K^{+} $-meson with the proton \textit{grows} with
energy the experimentalists, faithfully trusting what then guru taught, 
decided that this is but a transitory effect and finally the constant value will be revealed at higher energies. In two years the story had repeated at the CERN ISR , 
this time with the proton-proton cross-section and at energies five times higher .
Physicists were very much embarrassed. First of all those theorists 
who managed to persuade the community of the undeniable asymptotic constancy
of the total cross-sections. Their desperate attempts to save the asymptotic constancy failed one after another. For the sake of fairness we have to notice that 18 years since the Heisenberg paper two bold theorists T. T. Wu and H. Cheng
published a paper where they predicted, in the framework of some 
specific model,  the asymptotic growth of the total cross-sections
 $ \sim \ln^2\!s$ though their arguments differed from those by Heisenberg. 
The experimental discovery of the growth of
the total cross-sections was not the discovery of exactly 
logarithmic growth - other  options had (and have) enough room -
posed, howbeit,  a difficult general problem before the
theorists: why and how do cross-sections grow? What is the physical mechanism underlying this phenomenon? Certainly, answers of the type "they grow because $ \Delta =0.08 $ " is not what we mean. We will touch
this problem below. Now I am  to turn directly to the results from the LHC.
\section{LHC: Diffractive Products}
\subsection{ Integral Cross-Sections: Total, Elastic and Inelastic \cite{1}}
Since the operation of the ISR and till 2010 no new data on the total cross-sections in $ pp $ interactions were available. Due to known similarity of $ pp $ and $\bar{p} p $ we relied on the results in $\bar{p} p $ obtained in $S\bar{p} pS $ at CERN and $ TEVATRON $ at Fermilab. " For sake of poverty" the cosmic data were also in use in spite of huge errors. These partially surrogate data indicated that cross-sections continue to grow. So, generally it was not a great surprise when first LHC measurements revealed that $ \sigma_{tot} ^{pp}$ have grown at energies 
two orders of magnitude higher than those at the ISR though the growth in cross-sections was only slightly more than two times. Nonetheless, I would like to emphasize that this result cannot be considered as a routine one or so because even with account of the $\bar{p} p $ and cosmic ray data $ \sigma_{tot} ^{pp}$ could go quite in different ways, up to the flattening.
As $\sigma_{tot} = \sigma_{el} + \sigma_{inel}  $ the growth of the total cross-section does not necessary lead to the growth of both $\sigma_{el}  $ and $\sigma_{inel} $. So the discovery made by the TOTEM Collaboration  that both $\sigma_{el}  $ and $\sigma_{inel} $ (confirmed afterwards by ATLAS and, in regard to inelastic cross-sections, by the CMS and ALICE Collaborations) grow seems quite non-trivial. Moreover, not only 
$\sigma_{el}  $ grows absolutely but the ratio $ \frac{\sigma_{el}}{\sigma_{tot}} $ has also grown from 18 percent at the ISR to 27 percent at the LHC.
In simple words, the higher is the relative  velocity of colliding protons the less they "want" to give up their energy for inelastic processes. Naive view of the colliding balls implies the idea that with the growth of energy they interact at larger and larger distances. The results to be mentioned in the next Section seem to confirm this.
If this trend will persist till the full dominance of elastic scattering or it will finally stop or changed for the opposite? This is a very exciting and pressing problem.
\subsection{ Interaction Radius Grows with Energy \cite{2} }

 Experiments at the ISR revealed the appearance of the dip in 
 $ {d\sigma^{pp}}/{dt} $ at $ \approx $ 1.3 $GeV^{2}$ (the onset of it has been "felt" already at the Serpukhov synchrotron). The dip (minimum) is a typical diffractive feature inherent to all ondulatory phenomena due to interference. Its position is inversely proportional to the square of something which can be vaguely called  "interaction radius".
 TOTEM has found that the dip moved closer to zero and arrived to $ \approx $ 0.5  $GeV^{2}$ at 7 TeV. Thus the interaction radius grows with energy. We have to keep in mind that this is not equivalent to the growth of the cross-sections.
A similar  phenomenon was observed by TOTEM for the so-called forward slope whose growth also signals the growth of the interaction radius. And again both phenomena are not the same.
\subsection{$ d\sigma/dt $ at Small Transferred Momenta: Simple or Complicated? \cite{3} }

The simple exponential form $ \exp (Bt) $ for $ {d\sigma}/{dt} $ in many cases appeared quite adequate but already at lower energies it was noticed that more complicated form 
$ \exp ( Bt + Ct^{2}) $ is not ruled out if even not better. Such a conclusion has been strengthened by the observation made by TOTEM that at the level of 7 $ \sigma $ the t-distribution should be described by   $ \exp ( Bt + Ct^{2}) $ or even by $ \exp ( Bt + Ct^{2} +Dt^{3}) $. It is not easy to interpret such a finding in simple non-formal terms. Some people see in this just the nearby $ 2m_{\pi} $ threshold in the Pomeron trajectory , others guess "oscillations"...
We can only say at the moment that different forms shown above are related with different forms of the "profiles" in the impact parameter space.
E. g., dependent on parameters $ B,C,D  $ the average (transverse) interaction radius may appear larger or shorter. Small-$ t $ form of the cross-section certainly influences the extracted values of the total cross-sections. Unfortunately, all such arguments still aren't related with physical meaning of these complications.

\subsection{Scattering Phase: How to Retrieve It and Why ? \cite{4}}
Protons are electrically charged and so it is impossible to get rid of the ubiquitous photons both surrounding the colliding protons and exchanged between them. This leads to a specific  phenomenon of the Coulomb-nuclear interference: there exists the region of $ t $ where Coulomb and strong ("nuclear") interactions are comparable. At lower transfers the Coulomb part absolutely dominates, at higher ones it dies-off giving the floor to the strong interactions. In the narrow interval of transfers the interference term is directly related to the phases of both participants. As to the Coulomb phase (without strong interactions) it can be calculated theoretically. Another case is the strong interaction phase. At the moment no feasible model exists for it. Those which dominated the "market" during several decades should be ruled out as theoretically inconsistent.
So the measurements provided by the TOTEM Collaboration on this subject are very valuable and are directly related to the possibility of the extraction of the phase from the data (even if only in a limited interval of $ t $ and with a certain degree of model dependence).
Whether we can invent some other way (besides the use the CN interference) to extract the phase of the strong interaction amplitude remains a paramount challenge.
The information on the strong interaction phase is extremely important for our understanding of the spatial character of the high-energy collisions : are they more central or more peripheral? The knowledge of phase can also give us an estimate of the longitudinal extent of the interaction region and it seems to be \textit{very} long.

\subsection{ Inelastic Diffraction }
Some words about inelastic diffraction. Traditionally the efforts of experimentalists were concentrated on single (SD) and double diffraction (DD) processes. Why are we interested in them? For not so high missing masses we deal with \textit{diffractive dissociation} of the colliding protons and they constitute together with elastic scattering a quasi-closed class of "kin" processes. The studies of these processes can give information on the properties of Reggeons (including Pomerons), their structure and relatively low-energy interaction with protons. High-mass inelastic diffraction is related with high-energy interaction of Reggeons and can help to reveal their internal structure. Some problem is related to the definition of what we call diffraction dissociation and what the generic inelastic diffraction. 
One more distinctive class of inelastic diffraction is so called central diffraction when quasi-elastically scattered protons give some of their energy to production of final states with small rapidities in the c.m.s. frame. The latter can be related to some hard subprocesses, e.g. high $ E_{T} $ jets and this provides a kind of short-wave diagnostics of the Pomerons.
The TOTEM and CMS Collaborations measured both single and double diffractive processes\cite{5}, while for the study of the central diffraction the joint project of CMS and TOTEM  ( CT-PPS) has started \cite{6} and gives its first results.
Theoretical interpretation of these results is on the march.

\section{Where is QCD?}
Up to now all model predictions failed to describe the data on the differential cross-sections measured by TOTEM, especially near the dip and to the right of it.
However, the failure of models (some or even all of them ) is not a conceptual disaster : after some repair the models try to cope with data again and this process repeats itself already during several decades. 
Certainly we would like to have something more fundamental to compare with the data, something the failure or success of which  would mean much for our world view.
LHC experiments successfully test the Standard Model in its electroweak part
with such crucial results as the proof of the Higgs existence.
Strong interaction part of the Standard Model is embodied in quantum chromodynamics(QCD). Can we test QCD in diffraction experiments?
Everybody knows that QCD is a weakly coupled theory when applied to the processes characterized by short distances with all pleasures of perturbative calculations.
At which distances do occur diffraction processes? Alas, in this case the characteristic spatial scales, as was mentioned in the Introduction, are by no means  small. Even worse, they grow with energy. And we are standing before the great problems of QCD at large distances which are very far from being resolved.
As realists, we understand that it is quite silly to hope that QCD can give us some exact results on the scattering amplitude of such complicated composite objects like protons. But we can fairly hope to use QCD for obtaining at least a few important parameters, such as the intercepts and the slopes of leading Regge trajectories. This could help us to understand, both quantitatively and  qualitatively, the gross features of diffractive processes at high energies.

\section{Acknowledgements}
I am grateful to the Organizers of QUARKS-2016 for invitation and very good organization of the seminar.
I also believe that we, theorists, have to thank our fellow experimentalists for giving us so rich food for our minds.

\end{document}